\documentclass[conference]{IEEEtran}
\IEEEoverridecommandlockouts
\usepackage{cite}
\usepackage{amsmath,amssymb,amsfonts}
\usepackage{algorithmic}
\usepackage{graphicx}
\usepackage{textcomp}
\usepackage{xcolor}
\usepackage{tabularx}
\usepackage{float}
\def\BibTeX{{\rm B\kern-.05em{\sc i\kern-.025em b}\kern-.08em
    T\kern-.1667em\lower.7ex\hbox{E}\kern-.125emX}}
\begin{document}

\title{LEAN: Light and Efficient Audio Classification Network
\thanks{978-1-6654-7350-7/22/\$31.00 ©2022 IEEE}
}

\author{\IEEEauthorblockN{Shwetank Choudhary}
\IEEEauthorblockA{\textit{Samsung R\&D Institute} \\
Bangalore, India \\
sj.choudhary@samsung.com}
\and
\IEEEauthorblockN{CR Karthik}
\IEEEauthorblockA{\textit{Samsung R\&D Institute} \\
Bangalore, India \\
cr.karthik@samsung.com}
\and
\IEEEauthorblockN{Punuru Sri Lakshmi}
\IEEEauthorblockA{\textit{Samsung R\&D Institute} \\
Bangalore, India \\
srilakshmi.p@samsung.com}
\and
\IEEEauthorblockN{Sumit Kumar}
\IEEEauthorblockA{\textit{Samsung R\&D Institute} \\
Bangalore, India \\
sumit.kr@samsung.com}
}

\maketitle

\begin{abstract}
Over the past few years, audio classification task on large-scale dataset such as AudioSet has been an important research area. Several deeper Convolution-based Neural networks have shown compelling performance notably Vggish, YAMNet, and Pretrained Audio Neural Network (PANN). These models are available as pretrained architecture for transfer learning as well as specific audio task adoption. In this paper, we propose a lightweight on-device deep learning-based model for audio classification, LEAN. LEAN consists of a raw waveform-based temporal feature extractor called as Wave Encoder and logmel-based Pretrained YAMNet. We show that using a combination of trainable wave encoder, Pretrained YAMNet along with cross attention-based temporal realignment, results in competitive performance on downstream audio classification tasks with lesser memory footprints and hence making it suitable for resource constraints devices such as mobile, edge devices, etc . Our proposed system achieves on-device mean average precision(mAP) of .445 with a memory footprint of a mere 4.5MB on the FSD50K dataset which is an improvement of 22\% over baseline on-device mAP on same dataset.
\end{abstract}


\section{Introduction \& Related work}
In recent years, along with speech recognition, identifying sound types using raw signals has become an emerging research area. Typically, audio pattern recognition is the umbrella word that consists of many tasks such as audio tagging \cite{b1}, acoustic scene classification \cite{b2}, audio classification \cite{b3}, and sound event detection. Identifying the type of sound from real-time signal has several use cases such as hearing disability removal, accessibility, intelligent smartphones, improving gaming experience, etc. Deep learning-based methods are de facto advanced solutions \cite{b4},\cite{b5}, \cite{b6}, \cite{b7}, \cite{b8} to detect a sound event in the digital signal. The task of identifying the sound typically requires a sufficiently large dataset especially when the problem demands multi-class multi-label classification. Several datasets have been released in previous years namely Kaggle Free sound dataset \cite{b14}, UrbanSound8K \cite{b12}, and ESC50 \cite{b13}. One of the greatest breakthroughs in the audio classification task is the release of a large-scale audio dataset called AudioSet \cite{b10} which is a publicly available dataset having 500 hours of clips corresponding to 527 classes. Dataset is based on Youtube videos and clips. However, some of the clips have been deleted and hence putting a limitation on open usage. In the year 2020, one effort to release a large-scale open dataset has been done by the research community by releasing Free Sound Dataset (FSD50K) \cite{b9} with raw files available under an open license. Our present work is based on this FSD50K dataset.
Identifying sound types for the multi-class multi-label dataset is a complex task. Robust models for such types of tasks require deep and complex networks which are capable of capturing sound patterns. Taking inspiration from the vision-based task, CNNs \cite{b15},\cite{b16} have been extensively used by researchers to detect sound patterns \cite{b17},\cite{b18}, emotion \cite{b6}, and language detection \cite{b11},\cite{b19}. Such models typically take fixed input ranging from 1 second \cite{b7},\cite{b8} to 10 seconds \cite{b11},\cite{b19} and produce fixed size embedding \cite{b7},\cite{b8} which is used for the classification task. Several deep and complex models have also been designed \cite{b4},\cite{b7} which have achieved the state of the art performance on AudioSet. In a recent breakthrough, attention-based networks proposed by Bahdanau's \& Loung's \cite{b20},\cite{b22} and multi-attention-based transformers \cite{b21},\cite{b5} have been widely adopted for vision, and audio tasks and have shown significant performance gain. One such example is the Audio transformer \cite{b5} on FSD50K which achieved a mean average precision score (mAP) of .537. It consists of a stack of transformer blocks where each block is followed by average pooling except the last block. Similarly, transformer-based Patchout faSt Spectrogram
Transformer (PaSST) \cite{b31} tries to overcome the computational complexity of transformers by introducing a method called Patchout which is the current state of the art on FSD50K with .653 mAP. However, despite performing so well, transformers are costly both in memory and computation level and difficult to port on-device. For example in PaSST, various models discussed have parameters varying from 50M to 80M which translates to very large model size. PSLA\cite{b32} proposes a collection of training techniques such as ImageNet pretraining, balanced sampling, label enhancement, model aggregation, etc. to boost the model accuracy and achieve the best mAP of .5671 on FSD50K with ensemble models. PSLA uses several EfficientNet (B0, B2 with attention) based architectures for ensembling and their base model called EfficientNet-B0 with single-headed attention has 5.36M model parameters on the AudioSet dataset. BYOL-A\cite{b33} pretrains representations of the input sound invariant to audio data augmentations, which makes the learned representations robust to the perturbations of sounds and achieves the best mAP of .448 on FSD50K. Some shortcomings with the above-discussed work are increased model complexity, larger training time coupled with large model size which leads to increased computation and memory cost and thus makes them unsuitable for resource constraints environments such as mobile devices.
Our work tries to reflect on these issues as it is a simple reproducible pipeline that uses Pretrained Yamnet and trainable Wave Encoder with just 4.58M model parameters and quantized on-device model size is 4.5MB and hence suitable for on-device deployment. Also, our proposed model requires just 1 second of the audio frame to make predictions whereas several of these works \cite{b31},\cite{b32} take input frame duration ranging from 10 to 30 seconds and thus make our system computationally efficient due to reduced pre-processing time for creating spectrograms.
To the best of our knowledge, Temporal Knowledge distillation for on-Device audio classification \cite{b35} only works on FSD50K which focuses on developing a lightweight system for on-device deployment. This \cite{b35} work adopts Teacher-Student based architecture for knowledge transfer and utilizes a heavy transformer-based model as a teacher and on-device models of various architectures as a student. Since this \cite{b35} work provides on-device model performance, we baseline this work for comparing our model results. Among the several baseline models discussed in the work, Att-RNN achieves the best on-device mAP of .3471 whereas our model achieves an mAP of .445 on the same dataset which is a 22\% improvement over the baseline model.

Mostly frequency domain features such as log-mel and Short-time Fourier transform (STFT) based spectrogram have been common choices \cite{b16},\cite{b17},\cite{b18} as input for an audio classification task. However, PANN \cite{b4} model on Audioset has demonstrated that time-domain features can act as a supplement to Frequency domain features in improving the model performance. We also get the inspiration from PANN to consider both time and frequency domain features as input to our model.
In this work, we propose a novel lightweight network called as Light and Efficient Audio Classification Network(LEAN) for on-device audio classification. LEAN consists of waveform based temporal feature extractor called as Wave Encoder and logmel-based Pretrained YAMNet as a feature extractor and cross attention based temporal feature realignment scheme.
Our contributions, in this study, can be summarized in the following way:
\begin{itemize}
\item We propose a lightweight on-device novel network called as LEAN which takes temporal and frequency features as input.
\item We introduce temporal feature realignment scheme using pre-trained embeddings through cross attention which leads to improved performance with slight increase in training parameters.
\item We show that despite being lightweight in memory and smaller frame input of 1 second, we achieve competitive performance with .4677 as the best mAP on GPU and .445 on-device mAP with just 4.5MB model size(quantized) on FSD50K dataset.
\item We provide class level performance analysis  and impact by our pipeline.
\end{itemize}

\section{dataset}
For our system evaluation, we use Free Sound Dataset (FSD50K) \cite{b9}. It is an open dataset containing over 51,000 audio clips for sound events and has 200 classes drawn from the AudioSet Ontology \cite{b25}. The total duration of all clips for this dataset is 108 hours. Sound clips in the dataset are weakly labeled i.e. labeling of classes is done at clip level instead of frame level.
Clips are multi-labeled and range from 0.3-30seconds in duration. Audio files consist of sounds from humans, sounds of things, animals, natural sounds, musical instruments, and more. We adopt FSD50K dataset as it is freely available under the creative commons license and popularly available for benchmarking Sound Event Detection models. FSD50K dataset is divided into three parts namely training, validation, and evaluation. We use training and validation split for training, fine-tuning the model and evaluation split for showcasing our results. For benchmarking our model performance, we use the same metrics as mentioned in the baseline system \cite{b9} such as mAP(Mean Average Precision). Mean average precision is an approximation of the area under the class’s precision-recall curve, which is more informative of performance when dealing with imbalanced datasets such as AudioSet and FSD50K in comparison to the average area under the curve of the receiver operating characteristic curve \cite{b36}. In addition, we also calculate mAUC-PR (Mean Area Under the Curve for Precision-Recall) and sensitivity index (d-prime) as an additional metric. We also refer to the unofficial implementation of FSD50K baseline systems \cite{b29} for producing our results.

\section{methodology}\label{methodology}
This section describes methods applied for audio classification. Our end-to-end proposed architecture is shown in Fig.~\ref{fig:architecture}. Here, we begin by introducing preprocessing steps that were applied for audio feature generation. We then, describe our proposed Wave Encoder along with a pre-trained YAMNet-based model. We then discuss our proposed system for cross-attention-based feature realignment.

\subsection{Dataset Preprocessing}\label{AA}
We downsample all audio clips to 16k Hz using the Sound library \cite{b26}. We consider a 1-second audio patch as input for all our training and testing. Labels for each 1-second patch are the same as that of the complete audio clip. We repeat the data for those clips whose duration was less than 1 second to get a fixed 1-second duration of the sample required by our model.

Our model consists of a two-channel network and hence takes two different inputs: 1. Raw waveform 2. Log-mel Spectrogram. Raw waveform for the 1-second patch is directly considered as input to Wave Encoder. For creating the log-mel spectrogram, we use the same setup and parameters as used in YAMNet. YAMNet takes a log-mel spectrogram having the shape as (96,64) which is corresponding to 960ms audio data. In each patch, 64 being mel bins and 96 being time dimension. At broader level, parameters are : Window length = 25ms, hop length = 10ms, Number of Mel Bands = 64, Mel Frequency range = [125 Hz, 7500 Hz].

\subsection{Pretrained YAMNet as Feature Extractor}\label{pretrainedyamnet}
One of the channels (we call it as the right channel) in our proposed model is CNN based feature extractor which captures spatial features of the given audio frame. Among the available large-scale audio dataset-based pretrained models for transfer learning, we choose YAMNet \cite{b8} over Vggish \cite{b7} and PANN’s \cite{b4}. YAMNet is mobilenet based architecture with the adoption of depth and pointwise convolution. It is a lightweight pretrained model with nearly 3.7 million weights and trained on the AudioSet dataset and achieved .306 mAP on the evaluation set. Although both Vggish and PANN models are better in terms of the performance compared to YAMNet but they are heavy networks both in memory and computation and hence not suitable for resource constraints environments like mobile phones and edge devices.

\begin{figure*}[t]
\centering
\includegraphics[width=0.8\linewidth]{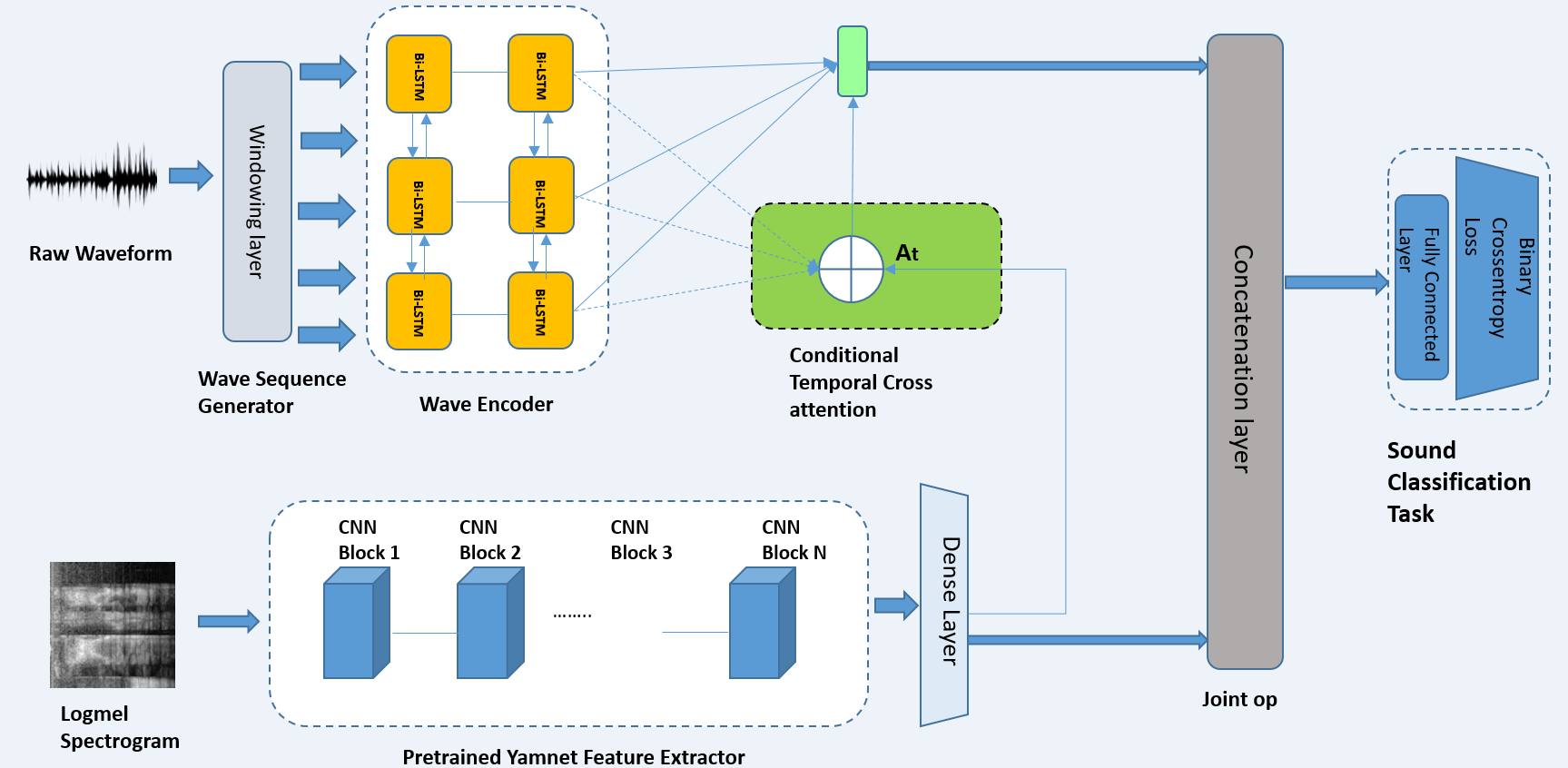}
\caption{Our Proposed Model Architecture (LEAN)}
\label{fig:architecture}
\end{figure*}

In our various experiments discussed below, we add a dense layer of 256 units (called as projection layer) after 1024 embedding vector of YAMNet as shown in  Fig.~\ref{fig:architecture}. The reason is to make adjustments concerning Wave Encoder output and to reduce joint embedding size.

\subsection{Proposed Wave Encoder as Temporal Feature Extractor}
Taking inspiration from the PANN model, we propose to use raw waveform input to capture time-domain features. In the PANN model, temporal features are captured by the Wavegram model which is a 1D Convolutional-based network. In contrast, the proposed Wave Encoder is a Bidirectional LSTM-based network. Wave Encoder takes raw waveform input and outputs learned temporal features. We call this part of the network as a left channel. It contains two Bi-directional LSTM layers each of 128 units. Since LSTM takes time sequence input, we first transform raw waveform into time sequence data by reshaping it. Reshape layer converts waveform into 2D time sequence data by splitting into patches using a non-overlapping window of 25 milliseconds. For 1 second frame with a sampling rate of 16K, reshape layer outputs a 2D vector as (40,400).

\subsection{Proposed Joint Model using Time and Frequency Domain features}
 We propose a joint model using Wave Encoder as the left channel and pretrained YAMNet with a projection layer as the right channel. Projection layer outputs embedding E\textsubscript{yam} = {e1,e2….,en}. Wave Encoder encodes raw waveform data into a context vector C\textsubscript{t}. Outputs from both the channels are combined using the following equation.
\begin{equation}
E_{combined} =  concatenate(E_{yam}, C_t)\label{eq1}
\end{equation}

If Y\textsubscript{i} is true multi label vector for i\textsuperscript{th} audio sample and \^{Y}\textsubscript{i} is predicted score vector, then our objective function can be defined as
\begin{equation}
\hat{Y_i}=sigmoid({E_{combined}}^T P+b)\label{eq2}
\end{equation}  
     
\begin{equation}
Loss =  BCE(Y_i, \hat{Y_i})\label{eq3}
\end{equation}  

\[Where \; E\textsubscript{combined} \;is  \;a  \;joint  \;embedding  \;vector  \;of  \;E\textsubscript{yam} \in R\textsuperscript{k} \]
\[and \; C\textsubscript{t} \in R\textsuperscript{k}  , E\textsubscript{combined} \in R\textsuperscript{k}  \;, P \in R\textsuperscript{k×c} ,\;b  \;as \; bias\] Here k is vector dimension and c is the number of classes. BCE is the binary cross entropy loss for multi label multi class problem.

During training, E\textsubscript{yam} is pretrained part and only C\textsubscript{t} part is learnt and updated during backpropogation.

\begin{table}[htbp]
\caption{Baseline, SOTA and proposed system performance on FSD50K Dataset}
\begin{center}

\begin{tabular}{|c|c|c|c|c|c|}
\hline
\textbf{Model Name} & \textbf{\textit{mAP}}& \textbf{\textit{mAUC}}& \textbf{\textit{dPrime}}& \textbf{\textit{\shortstack{Total\\ Param}}} \\
\hline
\shortstack{CRNN \cite{b9}\\(Baseline)}&.417 & - &2.068&0.96M  \\
\hline
\shortstack{VGG-like \cite{b9}\\(Baseline)}&.434&-&2.167 &0.27M  \\
\hline
\shortstack{ResNet-18\cite{b9}\\(Baseline)}&.373&-&1.883 &11.3M \\
\hline
\shortstack{DenseNet-121\cite{b9} \\ (Baseline)} &.425&- &2.112&12.5M  \\
\hline
\shortstack{Audio  \\ Transformer \cite{b5}} &.537&- &- &2.3M  \\
\hline
\shortstack{PSLA \cite{b32}} &0.5671&- &- & 13.64M \\
\hline
\shortstack{BYOL-A \cite{b33}} &0.448&0.896&- &6.3M  \\
\hline
\shortstack{PaSST \cite{b31} \\\textbf{(SOTA)}} &.653&- &-&50M  \\
\hline
\shortstack{Wav2CLIP \cite{b34}} &0.4308& - &- & $~$11M  \\
\hline
\shortstack{TKD \cite{b35}} &0.3471&-&- &0.2M  \\
\hline
\shortstack{\textbf{Joint Model  + Bahdanau} \\\textbf{Cross Attention}\\ \textbf{(LEAN)}} & \textbf{0.4677}&\textbf{0.944} &\textbf{2.251} &\textbf{4.58M}  \\
\hline
\end{tabular}
\label{tab1}
\end{center}
\end{table}

\subsection{Temporal feature realignment using Cross Attention for Joint Model}\label{crossattentionsection}
Instead of the simple concatenation of output from Wave encoder and YAMNet, we propose two realignment schemes using cross attention for getting final embeddings.
\begin{itemize}
\item Cross attention as simple affinity score
\item Cross attention similar to Bahdanau’s type attention with learning weights
\end{itemize}

\paragraph{Cross attention as simple affinity score }
Inspired by the work "Multimodal Speech Emotion Recognition Using Audio and Text" \cite{b6}, we adopt similar cross attention for our proposed lightweight system to improve the model performance without a significant increase in model parameters. The said work uses audio embeddings to realign textual embeddings. In a similar fashion, we propose to use pretrained YAMNet embeddings to realign Wave Encoder output using cross attention. In this scheme of attention, we use a simple dot product between YAMNet and Wave Encoder output to calculate affinity scores of time sequences from Wave Encoder.
Equations \eqref{eq4} to \eqref{eq7} discuss the mathematical formulation of the cross attention scheme. Here, if E\textsubscript{yam} is output of pretrained YAMNet embeddings after projection layer and h\textsubscript{t} is hidden state of t\textsuperscript{th} sequence of Wave Encoder output, then A\textsubscript{t} as attention vector is calculated using equation \eqref{eq4}. C\textsubscript{att} in equation \eqref{eq5} is the attentive context vector which is calculated using a weighted sum of T hidden states . C\textsubscript{att} is then concatenated with E\textsubscript{yam} to get final joint embeddings . Our objective function remains the same as discussed in equations \eqref{eq2} and \eqref{eq3}.

\begin{equation}
A_{t} = softmax(tanh(Dot(E_{yam}, h_{t})))\label{eq4}            
\end{equation}

\begin{equation}
C_{att} = \Sigma {A_t h_t}\label{eq5}         
\end{equation}

\begin{equation}
E_{combined(att)} =  concatenate(E_{yam}, C_{att})\label{eq6}                  
\end{equation}

\begin{equation}
\hat{Yi}= sigmoid({E_{combined(att)}}^T  P+b)\label{eq7}                  
\end{equation}

\paragraph{Cross attention similar to Bahdanau’s type attention with learning weights}
We experiment Bahdanau’s additive style of attention with learning weights using dense layers. We conceptualize cross attention as Query (Q), Key (K), and Value (V) in a similar fashion as done in the Transformer network \cite{b21}. In this scheme , projection layer output embeddings are chosen as Q, and Wave Encoder output is chosen as K, V.

The key difference in this method in comparison to affinity attention is in calculating A\textsubscript{t} attention vector which is discussed in equations \eqref{eq8} to \eqref{eq11}.
The below equations discuss the calculations involved in the cross-attention scheme.

\begin{equation}
Q = ({E_{yam}}^T W_{q} + b_{q})\label{eq8}                  
\end{equation}

\begin{equation}
K =  ({h}^T W_{k} + b_{k} )\label{eq9}                
\end{equation}

\begin{equation}
V = (tanh(Q +K) W_{v} + b )\label{eq10}                                       
\end{equation}

\begin{equation}
A_{t} = softmax(V)\label{eq11}      
\end{equation}

 \[Where \; E\textsubscript{yam}  \in \mathbb{R} \textsuperscript{(m)}   \; , \;W\textsubscript{q}  \in \mathbb{R}\textsuperscript{m×d},  \; h  \in \mathbb{R}\textsuperscript{(t×m )}\] 
\[ W\textsubscript{k}   \in \mathbb{R}\textsuperscript{m×d}, W\textsubscript{v}   \in \mathbb{R}\textsubscript{m×1} \] 
Here d is dense layer units and is choosen as 128  after several experiments.

\begin{table*}[htbp]
\caption{Class wise mAP improvement using proposed model on 200 classes for FSD50K}
\begin{center}
\begin{tabular}{|c|c|c|c|c|c|}
\hline
\multicolumn{2}{|c|}{\textbf{ \textit{Baseline\& Reference Models}}}&\multicolumn{4}{|c|}{\textbf{ \textit{Count of classes which see improvement(mAP) over B by R out of 200 classes}}} \\
\hline
\textbf{ \textit{Baseline Model(B)}} & \textbf{\textit{Reference Model(R)}}& \textbf{\textit{Overall Improvement}}& \textbf{\textit{Better by 5\%}}& \textbf{\textit{Better by 10\%}}& \textbf{\textit{\shortstack{Better by 20\%}}} \\
\hline
YAMNet Finetune & Joint Model & 140 &  55 & 31& 13  \\
\hline
YAMNet Finetune & Joint Model with Cross Attention & 135 & 53 & 30 & 14  \\
\hline
Joint Model &  Joint Model with Cross Attention & 96 & 27 & 9 & 3  \\
\hline
\end{tabular}
\label{tab:classwiseap}
\end{center}
\end{table*}

\subsection{Training and Hyper Parameter Selection}
We use Tensorflow-GPU 2.3.0 \cite{b27} and Keras 2.6.0 \cite{b28} for all our implementation. For all our training, we fix batch size as 64, learning rate as 1e-4, and loss as binary cross-entropy.

Training is done for 40 epochs with each epoch roughly taking 1.5 hours. We observe that models generally converge well before the last epoch. We select the best model having the highest validation AUC (PR).

\section{RESULTS \& DISCUSSION}

We use Nvidia GPU GeForce GTX 1080 Ti 11178 MB card for training and testing. All models are trained and tested end to end. Testing is done on the evaluation dataset provided by FSD50K. For testing, a given audio file is split into chunks of 1 second each with an overlap of 50\% . For each chunk, the raw waveform and corresponding log-mel (with mean normalization) are calculated and fed to the network. Chunk level class predictions score is calculated and averaged to get a final class-wise score for the entire file. We observe a better performance with an overlapping split compared to the non-overlapping one. We believe this is mainly due to the availability of more data by the overlapping split.

Below, we discuss all results and the type of attention network we adopt.

\subsection{Fine Tuning YAMNet}
As discussed in section \ref{pretrainedyamnet}, we choose YAMNet as our feature extractor which gives 1024 vectors as latent embedding. We experiment in multiple ways to fine-tune YAMNet to achieve its best performance on FSD50K. For this, we retrain YAMNet end to end with and without pretrained weights initialization. However, we observe that none could outperform the baseline system \cite{b9}. We then directly use pretrained embedding with a projection dense layer of 256 and achieved an mAP of 0.4577 outperforming baseline results of VGG-like having mAP of .434.

\begin{table}[htbp]
\caption{Ablation Study}
\begin{center}
\begin{tabular}{|c|c|c|c|c|c|}
\hline
 \textbf{\textit{Model}}  & \textbf{\textit{\shortstack{mAP}}}& \textbf{\textit{mAUC}}& \textbf{\textit{dPrime}}& \textbf{\textit{\shortstack{Total \\ Param}}} \\
\hline
\shortstack{YAMNet \\ (without \\  finetune)} &0.446&0.939&2.192 &3.42M  \\
\hline
\shortstack{YAMNet \\ (finetune)} &0.4577&0.943 &2.235&3.53M  \\
\hline
\shortstack{Joint Model(JM)} &0.4609&0.944 &2.243 &4.51M  \\
\hline
\shortstack{JM + Affinity\\ Cross Attention} &0.4654&0.944&2.249&4.51M  \\
\hline
\shortstack{ \textbf{JM  + Bahdanau} \\ \textbf{Cross Attention}\\  \textbf{(LEAN)} } & \textbf{0.4677}& \textbf{0.944} &\textbf{2.251}& \textbf{4.58M}  \\
\hline
\end{tabular}
\label{tab:ablation}
\end{center}
\end{table}

\begin{table}[htbp]
\caption{on-Device Performance comparison on FSD50K dataset}
\begin{center}
\begin{tabular}{|c|c|c|c|}
\hline
\textbf{\textit{Model}} & \textbf{\textit{Model Size}}& \textbf{\textit{mAP}}& \textbf{\textit{Inference}}  \\
\hline
\textbf{LEAN} & \textbf{4.5 MB} & \textbf{0.445}  & \textbf{65ms}   \\
\hline
Att-RNN \cite{b35} & - &0.3471&-  \\
\hline
MHAtt-RNN \cite{b35} & - & 0.3317&-  \\
\hline
CRNN \cite{b35} & - &0.3053& -  \\
\hline
\end{tabular}
\label{tab:ondevice}
\end{center}
\end{table}

\subsection{Temporal feature extractor as Wave Encoder}
In the Wave Encoder, we use two Bi-LSTM layers each of 128 units. Context vector yielded from wave encoder is concatenated using equation \eqref{eq1} and trained using objective function discussed in equations \eqref{eq2} and \eqref{eq3}. Using this configuration, we get the mAP of 0.4609. We call this model scheme as Joint Model(JM).

As shown in Table \ref{tab:ablation}, our simple joint fusion system (YAMNet + Wave Encoder) surpasses fine-tuned YAMNet in all performance metrics.

\subsection{Variation of cross attention and their impact}

Table \ref{tab:ablation} captures our various architechture experiments as ablation study. In our first cross attention scheme, we consider an affinity-based attention model for the realignment of temporal embeddings and achieve an mAP of 0.4654. We call this model a "Joint model(JM) with affinity Cross attention". From Table \ref{tab:ablation}, we observe that both mAP and dPrime see an improvement compared to fine-tuned YAMNet and Joint model. Another observation is that this scheme of attention does not add any training parameters but still leads to improvement in results which highlights the novelty of this approach. We believe that reason for this improvement due to realignment is the identification of relevant temporal sequences using attentive weights. This mechanism gives higher weights to relevant features and the rest are suppressed by low weighing.

In our next experiment, we adopt Query (Q), Key (K), and Value (V) type structure as discussed in section \ref{crossattentionsection} using Bahdanau \cite{b20} additive style of attention. We call this scheme as our proposed LEAN model. Based on our experiments and to keep the model size small, we choose 128 dense units for calculating Q, K, and V. With this setup, we achieve our model best mAP of .4677 and dPrime value of 2.251.

\subsection{Class level performance analysis}
Since the baseline system does not contain class-wise results, we compare the class-level performance of our own fine-tuned YAMNet model with the proposed LEAN model.

Table \ref{tab:classwiseap} shows class-wise mAP analysis of different models. We use Baseline(B) and Reference Model (R) nomenclature for convenience where B is compared with R. In the table, the "Overall Improvement" column tells about the number of classes in R that see improvement in mAP over B.

We observe that by introducing a wave encoder, we see an improvement in 70\% of the classes compare to YAMNet and further improvement is seen in 48\% classes by the realignment scheme.

\subsection{on-Device experiments}
Our on-device model results are discussed in Table \ref{tab:ondevice}. For deploying on-device, we perform quantization using TensorFlow tflite which results in a model size of nearly 4.5MB. We implement pre-processing steps in the same manner as discussed in section \ref{methodology}. We use the Samsung S21 smartphone (Android SDK 30, 12GB RAM, 256GB ROM, Octa-core, Exynos2100 chip) for our on-device experiments. Due to model quantization, we see slight natural drop in mAP from .4677 on GPU to .445 on on-device. We compare our on-device model result with this \cite{b35} baseline work which demostrates several on-device models with best mAP by Att-RNN model as .3471 which is 22\% lower than our LEAN model.
\section{CONCLUSION}
In this paper, we propose a lightweight joint embedding-based audio classification model for on-device deployment called as LEAN which simultaneously extracts temporal and pretrained spatial features from the given audio signal. Further, we use novel temporal features realignment using cross attention which results in improved performance with a slight increase in model parameters. Our proposed LEAN model outperforms the baseline on-device system on the FSD50K dataset with a lesser memory footprint and produced competitive results in comparison to several existing works. We also conduct a detailed analysis of class-level impact using our system for all 200 classes of FSD50K dataset. Our work demonstrates that using a pretrained model, it is possible to achieve further performance improvement using cross-attention-based temporal realignment. In the future work, we aim to replace YAMNet with state of an art pretrained transformer-based feature extractor model and see the performance impact. In addition to this, we wish to leverage label ontology-based mutual relation and dependency knowledge for FSD50K multi-label dataset to improve class-level performance.

\end{document}